\pdfoutput=1
    \documentclass[%
     reprint,superscriptaddress,
     amsmath,amssymb,
     aps,
    prx,
    ]{revtex4-2}
    
    \usepackage{graphicx}
    \usepackage{dcolumn}
    \usepackage{bm}
    \usepackage{bbm}
    \usepackage{float}
    \usepackage{soul}
    \usepackage{color}
    \usepackage{siunitx}
    \usepackage{hyperref}
    \usepackage{amsmath}
    
    \makeatletter
    
    \newcommand{\hypgeo}[2]{%
      {\vphantom{F}}_{#1}\kern-\scriptspace F_{#2}%
    }
    
    \newcommand\approxsim{\mathchoice
      {\@approxsim {\displaystyle}      {1ex} }
      {\@approxsim {\textstyle}         {1ex} }
      {\@approxsim {\scriptstyle}       {.7ex}}
      {\@approxsim {\scriptscriptstyle} {.5ex}}}
    \newcommand\@approxsim[2]{%
      \mathrel{%
        \ooalign{%
          $\m@th#1\sim$\cr
          \hidewidth$\m@th#1.$\hidewidth\cr
          \hidewidth\raise #2 \hbox{$\m@th#1.$}\hidewidth\cr
        }%
      }%
    }

    \newcommand{\bo}{\raise-1mm\hbox{\Large$\Box$}}

    \newcommand{\ed}[1]{{\color{black}{#1}}}

    \renewcommand{\appendix}{
        \par
        \setcounter{section}{0}
        \renewcommand{\thesection}{\Alph{section}}
        \renewcommand{\thesubsection}{\thesection.\arabic{subsection}}
        \setcounter{page}{1} 
        \setcounter{equation}{0} 
        \renewcommand{\theequation}{S\arabic{equation}} 
    }
    
\begin{document}
    
    \preprint{APS/123-QED}
    
    \title{\ed{Counting statistics and inference limits for reactions driven by latent ergodic dynamics}}
    
    \author{Benjamin Garcia de Figueiredo}
    \affiliation{ICTP South American Institute for Fundamental Research \& Instituto de F\'isica Te\'orica, Universidade
    Estadual Paulista - UNESP, R. Dr. Bento Teobaldo Ferraz, 271 - 2 - Várzea da Barra Funda, São Paulo - SP, 01140-070, Brazil}
    \affiliation{Lewis–Sigler Institute for Integrative Genomics,
    Princeton University, Princeton, NJ 08544 USA}
    \author{Justin M. Calabrese}
    \affiliation{Center for Advanced Systems Understanding (CASUS) -- Helmholtz-Zentrum Dresden-Rossendorf (HZDR), Untermarkt 20, Görlitz 02826, Germany}
    \affiliation{Department of Ecological Modelling, Helmholtz Centre for Environmental Research – UFZ, Leipzig, Germany}
    \affiliation{Department of Geosciences, TUD Dresden University of Technology, Dresden, Germany}
    \author{William F. Fagan}
    \affiliation{Dept. of Biology, University of Maryland, College Park MD 20742, USA}
    \author{Ricardo Martinez-Garcia}%
     \email{r.martinez-garcia@hzdr.de}
    \affiliation{Center for Advanced Systems Understanding (CASUS) -- Helmholtz-Zentrum Dresden-Rossendorf (HZDR), Untermarkt 20, Görlitz 02826, Germany}
    \affiliation{ICTP South American Institute for Fundamental Research \& Instituto de F\'isica Te\'orica, Universidade
    Estadual Paulista - UNESP, R. Dr. Bento Teobaldo Ferraz, 271 - 2 - Várzea da Barra Funda, São Paulo - SP, 01140-070, Brazil}
    \affiliation{Department of Ecology, Institute of Biosciences, University of São Paulo, São Paulo, Brazil}
    
    \date{\today}
    
    \begin{abstract}
    \ed{Many natural phenomena--from chemical reactions to biological encounters and sensor detections--are described through counts of discrete events. Such events occur at random times when an underlying, often unobserved, dynamical system enters a reactive region of its phase space. Despite the ubiquity of these processes in nature, a general framework linking counting statistics to the properties of the underlying stochastic process is lacking. Most existing theories, including widely used mass-action models, neglect trajectory correlations and assume independent counts. We develop a theory that overcomes these limitations, deriving a general expression for the inter-count time distribution as a function of the spatiotemporal scales of the reactive region and the stochastic process. The counting statistics interpolate between a reaction-limited regime, where counts are effectively independent and follow Poisson statistics, and a diffusion-limited regime, where counts are strongly determined by the correlation structure of the latent stochastic process and exhibit systematic overdispersion relative to the Poissonian limit. This structure establishes fundamental constraints on what properties of the stochastic process can be inferred from measured event counts alone. We illustrate these constraints in two biophysical examples: an activity-monitoring assay for a small organism based on detector crossings and population size estimation from encounter counts, in which we recover and generalize classical sensing limits. Our framework thus provides a unified theoretical description of counting processes driven by latent diffusion dynamics, and quantifies how much information each process carries about the other.}
    \end{abstract}
    
    \maketitle
    

    \section{Introduction}\label{sec:intro}
    Many natural phenomena, such as radioactive decay, cell division, neuronal spiking, or disease transmission, can be described as counting processes in which a set of discrete random variables $N(t)$ change at randomly distributed time intervals as they accumulate the number of observed events \cite{VanKampen1992}. In many of these examples, the measured events are triggered when an underlying dynamical variable $\bm z(t)$ enters a subset of its phase space, $\Omega$, within which reactions occur at some rate $\nu$. Under these conditions, $N(t)$ is driven by a stochastic process in a latent phase space (Fig.\,\ref{fig:mapping}A). 
    
    Counting processes driven by latent stochastic dynamics are common in natural systems, particularly in biological contexts. In cell biology, for example, the transport of molecules between cellular compartments controls many intracellular reactions. For example, the binding of transcription factors (TF) to DNA \cite{wang_quantitative_2009} depends on TF diffusion from the cytoplasm to the chromosome. Thus, the number of binding events $N(t)$ associated to a TF-binding domain pair will be proportional to the occupation time of the binding domain volume $\Omega$ by the TF trajectory $\bm z(t)$, times a binding rate $\nu$. 
    Trajectory encounter counts also underlie many ecological interactions, such as disease transmission \cite{kenkre_theory_2014, sugaya_analysis_2018}, predation \cite{miller_estimating_2013, scrafford_temporal_2018, coblentz_estimating_2021}, human-wildlife conflict \cite{hels2001,grilo2018}, or pollination \cite{bosch2009,cole_first_2017}. Other examples come from animal behavior, where experimental protocols often involve inferring activity levels from sensor counts \cite{pfeiffenberger_locomotor_2010, rowcliffe_quantifying_2014}, and switches between behavior states can be analyzed as a drift-diffusion process in a latent space of neuronal activity \cite{gold_neural_2007}.
    
    \ed{A key property of many stochastic dynamics, even when not directly observed, is the existence of a stationary distribution. Physical and biological constraints often confine the dynamics, so that the probability density of $\bm z(t)$ converges to a limiting distribution $p_{\mathrm{st}}(\bm z)$ as $t\rightarrow \infty$. When initial conditions are unknown, this stationary measure also describes the distribution of independent realizations, making time and ensemble averages coincide. This property is known as ergodicity, and it plays a key role in linking trajectory to counting statistics.}
    
    In this manuscript, we derive how the correlation structure of the trajectory $\bm z(t)$ affects the statistics of the count $N(t)$ under the assumption of ergodicity. In the context of elementary reaction kinetics and population dynamics, the first approximation is to ignore any inner structure of $\bm z(t)$ in favor of the law of mass action. This model assumes that $N(t)$ increases proportionally to the rate $\nu$ and the concentration of reactive pairs, as the distribution $\bm z(t)$ very quickly approaches a uniform profile. This assumption can be refined by making the reaction rate dependent on space to account for inhomogeneities in spatial distribution \cite{MartinezGarcia2020}. In either case, trajectories are subsumed under a stationary density, and the process is said to be reaction-limited. However, even in the simplest case of molecules diffusing by Brownian motion, deviations from stationarity can play an important role. In these scenarios, a full description of the trajectories $z(t)$ is necessary to describe the counting statistics. 
    
    A classic example where deviations from the stationary limit are essential is the Berg–Purcell chemosensing limit \cite{berg_physics_1977}, which shows that the correlation structure of molecular trajectories sets how well a cell can sense a chemical gradient through membrane binding. Because single molecular trajectories dwell around their instantaneous positions, binding events occur in correlated bursts rather than independently, and this behavior can be studied by computing how often a molecule approaches or moves away from the vicinity of the cell. This example motivates investigating another extreme case, in which the target's reaction rate is mostly set by the time $T$ it takes a trajectory to reach $\Omega$ from its initial position, which can be analyzed using first-passage theory \cite{redner_guide_2001}. Both approaches can be very accurate but lack generality because they require a large separation of time scales between $1/\nu$ and $T$, and the latter approach also requires additional assumptions about how the trajectory is stopped or reset after each reaction. 
    
    The dichotomy between reaction- versus diffusion-limited processes has been studied many times within the reaction-diffusion literature \cite{kenkre_investigation_1985}. Recently, this distinction has ignited debate in the field of movement ecology \cite{das_misconceptions_2023} about which approach is appropriate for animal movements studies, in which trajectories can have non-trivial correlation structure and reactions (e.g., encounters between individuals) are sparsely observed. In some contexts, one can combine both approaches and compute the statistics of the first reaction time $R$, which rigorously interpolates the hitting time $T$ and a stationary time scale proportional to $1/\nu$ \cite{bicout_first_1997}. Still, $R$ alone does not directly reflect what information about trajectories remains relevant for $N(t)$ as $\bm z(t)$ approaches its stationary limit, because it only accounts for a single reaction. As in the example of the Berg-Purcell limit, the count $N(t)$ reflects correlations within a persistent trajectory. Therefore, the study of the statistical structure of $N(t)$ in the literature has so far remained restricted to a few specific, mathematically tractable processes, whereas modern tools in biology increasingly enable the reconstruction of trajectories with complex statistical properties.
    
    \ed{We address this gap in two steps. First, we develop a general theory to characterize reaction counts $N(t)$ assuming only that $\bm z(t)$ is an ergodic diffusion and $\Omega$ is small relative to the length scales of motion (section \ref{sec:exact}). We then illustrate how this framework informs the interpretation of counting data in two biological scenarios \ed{(section \ref{sec:counting})}. In particular, we focus on how much information each process carries about the other, quantified by how much of the statistics of one can be inferred from observations of the other. We conclude by discussing broader implications, limitations and possible extensions of the theory (section \ref{sec:conclusion})}
    
    \section{Results}
    \subsection{Problem setup and main results}\label{sec:prem}
    To specify $N(t)$ as a counting process, it suffices to specify the distribution of the time intervals between two consecutive counts. When these times are independent and identically distributed, they are called holding or sojourn times. If they are additionally distributed such that the mean time between consecutive reactions is finite, the counting process is called a renewal process \cite{smith1958renewal, daley_introduction_2008}. 
    If counted events are themselves independent, the time intervals between consecutive events become exponentially distributed and $N(t)$ becomes a Poisson random variable. When $N(t)$ is a Poisson process, reaction rates straightforwardly map to differential equations for moments and efficient simulation schemes for stochastic dynamics \cite{VanKampen1992, Gillespie1977, Gibson2000, Tian2004, Chatterjee2005}. While these are useful results, in general ignoring the non-exponential nature of holding times can introduce significant errors in the analysis of the counting process \cite{Castro2018,coblentz_estimating_2021}, which makes understanding the correlation structure of counts important.
    
\begin{figure*}
    \centering
    \includegraphics[width=0.75\textwidth]{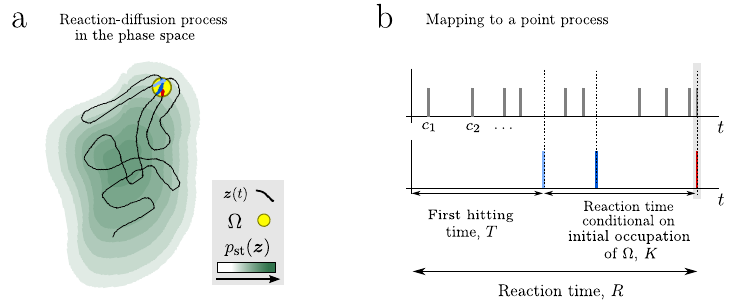}
    \caption{A) Schematic of the diffusion process, represented by its stationary distribution $p_\mathrm{st}$ (green shaded region) and a trajectory (black line), and a reactive domain $\Omega$ (yellow area). B) Representation of the reaction-diffusion process in terms of a point process. The top panel represents a possible sequence of potential interaction times, $C_\Omega= \lbrace c_i \rbrace$, and the bottom panel represents the sequence of times at which the diffusion process crosses the reaction domain. The last crossing event (red line and shaded in gray) represents a reaction event because it coincides with one of the times in $C_\Omega$}\label{fig:mapping}
    \end{figure*}

    Because the occupation-conditioned reaction rate $\nu$ is still assumed constant, the intervals between consecutive counts of $N(t)$ come from a Poisson process that is defined not in the experimental time $t$ but in the occupation time of $\bm z(t)$ in $\Omega$, 
    \begin{equation}\label{eq:occtime}
    \ell_\Omega(t) = \int_0^t \bm 1[\bm z(t) \in \Omega]\mathrm{d}t.
    \end{equation} 
    where $\bm 1$ is the indicator function. In other words, there is a Poisson point process of times at a rate $\nu$, given by a sequence $(c_\alpha)_{\alpha= 1}^\infty$, such that a reaction happens at time $c_\alpha$ if $\bm z(c_\alpha) \in \Omega$ (Fig.\,\ref{fig:mapping}B). The sequence of intervals $(K_i)_{i =1}^\infty$, between consecutive times that meet this condition specifies $N(t)$. One reason for the ubiquity of the Poisson law for $N(t)$ is therefore the law of rare events: If the inter-reaction interval $K_i$ is larger than the typical decorrelation time of $\bm z(t)$, reaction events are approximately independent. 
    
    \ed{In this work, we assume that $\bm z(t)$ is a Markov process, such that its future values depend only on its present state and not on its past history. We model $\bm z(t)$ by the It\^o stochastic differential equation
    \begin{equation}\label{eq:langevin}
        \dot{{\bm z}} = \bm \mu(\bm z) + \Sigma(\bm z)\, \bm \xi(t),
    \end{equation}
    where ${\bm \xi}$ is a Gaussian white noise process with zero mean and identity covariance matrix modulated by the instantaneous covariance matrix $\Sigma(\bm z)$, and $\bm\mu(\bm z)$ is the drift function.
    
    We further assume the process is ergodic, so it admits a unique stationary probability distribution $p_\mathrm{st}$ to which it converges in the long term starting from any initial condition. Equivalently, time averages along almost every trajectory equal ensemble averages under $p_\mathrm{st}$. In particular, for any compact region $\Omega$, the first-passage hitting time \cite{redner_guide_2001}
    \begin{equation}
        T = \min \{ t \in [0,\infty) \mid \bm z(t) \in \Omega \}
    \end{equation}
    has a finite expected value that defines the characteristic time scale at which $\bm z(t)$ loses its correlations.}
    
    Because $\bm z(t)$ is an ergodic Markov process, the time intervals between two consecutive reactions, $(K_i)_{i=1}^\infty$, are approximately independent and have a finite expectation, making $N(t)$ a renewal process where correlations between consecutive reactions are set by the time scales of the distribution of $T$. If $1/\nu$ is large compared to these correlation time scales, the motion is being sampled in a regime of ergodic averaging, so $\ell_\Omega(t) \approx t \int_\Omega p_\mathrm{st}(\bm z)\mathrm{d}\bm z$, and $N(t)$ is approximately Poisson with a rate $\omega_\mathrm{st} = \nu\int_\Omega p_\text{st}(\bm z)\mathrm{d}\bm z$. In any other case, the inter-reaction times are limited by $T$ and as we will derive, follow a distribution given implicitly by,
    \begin{equation}\label{eq:Kmgf0}
        \langle e^{-sK_i} \rangle = \frac{1}{1 + \frac{s}{\omega_\mathrm{st}}\langle e^{-sT}\rangle},
    \end{equation}
    where $\langle \cdot \rangle$ denotes an expectation where the initial conditions $\bm z(0)$ are sampled from $p_\mathrm{st}$. This distinction of regimes set by the scales of $\nu$ and $T$ has been previously explored at the level of averages \cite{bicout_first_1997}, i.e. to the first order in $s$ of Eq. \eqref{eq:Kmgf0}. In applications where the object of interest is the first reaction time $R = T + K$, where $K$ follows the distribution of $K_i$, reactions are said to be diffusion-limited when $\langle R \rangle \approx \langle T \rangle$ and reaction-limited when $\langle R \rangle \approx \langle K \rangle \propto 1/\nu$ \cite{das_misconceptions_2023}. In the next sections, we present a new framework to obtain the full statistics of $R$ when this separation of time scales is not evident. In doing so, we also recover the statistics of $K$ by considering $R$ conditioned on initialization at $\Omega$, which fully characterizes the counting statistics.
    \subsection{Characterizing the next-reaction time distribution}\label{sec:exact}
    The statistics of $R$ can be obtained by solving for the probability density of the process $\bm z(t)$ stopped at time $R$. We denote by $P(\bm z,t|\bm z_0)$ the propagator for this process, which is the probability density function of the motion given a deterministic initial condition $P(\bm z, 0) = \delta(\bm z-\bm z_0)$. Because the local reaction is a Poisson process, it can be written simply as a constant sink for probability conditioned on the occupation of $\Omega$. We can therefore write down the backward Fokker-Planck equation \cite{gardiner_stochastic_2009} for the process as
    \begin{align}
    \label{eq:FPEback}
    \frac{\partial P(\bm z, t|\bm z_0)}{\partial t} = \left[\hat L^\dagger_{\bm z_0} - \omega(\bm z_0)\right]P(\bm z,t|\bm z_0),
    \end{align}
    where $\hat L^\dagger_{\bm z_0}$ is the adjoint of the Fokker-Planck operator (also called the generator of diffusion) for Eq.\,\eqref{eq:langevin},
    \begin{equation}\label{eq:fpoperator}
        \hat L^\dagger = \bm \mu \cdot \bm \nabla + \frac{\Sigma \Sigma^\intercal}{2}:\bm{\nabla} \bm{\nabla},
    \end{equation}
    and the function $\omega$, also called the density of the killing measure \cite{borodin_handbook_2002}, is the space-dependent reaction rate. We take the statement of $\Omega$ being small as meaning that there exists a point $\bm z_\Omega$ such that
    \begin{equation}\label{eq:omegadef}
        \omega(\bm z) = \nu\bm 1[\bm z \in \Omega] \approx \eta \delta(\bm z - \bm z_\Omega),
    \end{equation}
    where $\bm 1$ is the indicator function and the variable $\eta = \nu\,\mathrm{vol}(\Omega)$ has dimensions of velocity in $1D$ and more generally of $\text{length}^d/\text{time}$ in $d$ dimensions.\\
    
    \textbf{The defect technique and local-global correspondence.} Eq. \eqref{eq:FPEback} implies an equation for the moment generating function (MGF) of $R$, $\langle e^{-sR}\rangle_{\bm z_0}$, shown in App.\ref{app:MGFderivation},
    \begin{equation}\label{eq:MGF}
        \left[\hat L^\dagger_{\bm z_0} - s - \omega(\bm z_0)\right]\langle e^{-sR}\rangle_{\bm z_0} = -\omega(\bm z_0),
    \end{equation}
    where we denote by $\langle \cdot \rangle_{\bm z_0}$ the expectation conditioned on a deterministic initialization of $\bm z$ at $\bm z_0$, and drop the subscript when conditioning on the stationary probability distribution $p_\mathrm{st}$. 
    This equation can be solved using the defect technique  \cite{Montroll1955, grosche_-function_1993, cavalcanti_exact_1999, kenkre_montroll_2021, kay_defect_2022}, which consists of formally inverting the equation in terms of a linear system for the Green's function $G_s(\bm z,\bm \xi)$ of the operator $[\hat L^\dagger - s]$ (App.\,\ref{app:MGFderivation}). This procedure gives
    \begin{align}\label{eq:MGFR}
        \langle e^{-sR} \rangle_{\bm{z}_0} &= -\frac{ G_s(\bm{z}_0,\bm{z}_\Omega)}{\frac{1}{\eta} - G_s(\bm{z}_\Omega,\bm{z}_\Omega)}.
    \end{align}
    indicating that the statistics of $R$ are fully determined by $G_s$. We now show how ergodicity constrains the structure of this Green's function, or equivalently, of the operator it inverts. This calculation is sensitive to dimensionality so we will consider a one-dimensional system for which our arguments are exact. The stationary distribution of a one-dimensional diffusion is, up to normalization,
    \begin{equation}
        p_{\text{st}}(z) \propto \frac{1}{\Sigma(z)^2}\exp\left({\int^z \frac{\mu(y)}{2 \Sigma(y)^2}\rm{d}y}\right),
    \end{equation}
    which allows the Fokker-Planck operator to be rewritten in canonical Sturm-Liouville form as
    \begin{equation}\label{eq:stationaryoperator}
        p_\text{st}(z)\hat{L}^\dagger_z = \frac{1}{2}\frac{\mathrm{d}}{\mathrm{d}z}\left(\Sigma(z)^2p_{\text{st}}(z)\frac{\mathrm{d}}{\mathrm{d}z}\right).
    \end{equation}
    This relation defines a so-called local-global correspondence between observables close to the boundary of $\Omega$ and their value initialized at equilibrium \cite{pitman_lengths_1997,bicout_first_1997,pitman_hitting_2003}: if the operator on the left hand side of Eq.\,\eqref{eq:stationaryoperator} is applied to a function and integrated over $z$, all contributions come from derivatives evaluated at the boundary of $\Omega$.
    
    To understand how ergodicity constrains the structure of the Green's function in Eq.\,\eqref{eq:MGF} and consequently of the MGF of $R$, we multiply both sides in \eqref{eq:MGF} by $p_\mathrm{st}(z_0)$ and take the limit $\nu \rightarrow +\infty$. In this limit, $R \rightarrow T$ and the killing measure can be swapped for Dirichlet boundary conditions $\langle e^{-sT}\rangle_{z_\Omega} = 1$. Assuming either a boundary at infinity or no-flux compact boundaries, we integrate over $z_0$, and the local-global correspondence, in terms of the MGF, takes the form
    \begin{equation}\label{eq:localglobal}
        -\frac{\Sigma(z_\Omega)^2p_\text{st}(z_\Omega)}{2}\frac{\mathrm{d}\langle e^{-sT}\rangle_{z_0}}{\mathrm{d}z_0}\bigg |^{z_\Omega^+}_{z_\Omega^-} = s \langle e^{-sT} \rangle,
    \end{equation}
    where, as introduced above, the absence of a subscript in $\langle e^{-sT} \rangle$ indicates that $T$ is conditional on an initial condition sampled from the stationary distribution of $z(t)$, $p_\mathrm{st}$. 
    
    Crucially, again due to one-dimensionality, the function $\langle e^{-sT}\rangle_{z_0}$ is monotonic on the half-intervals to either side of $z_\Omega$, where it attains a maximum of unity. Thus, the functional form of $\langle e^{-sT}\rangle_{z_0}$ restricted to $z_0 < z_\Omega$ or $z_0 > z_\Omega$ gives the two linearly independent solutions to the homogeneous problem for $[\hat L^\dagger - s]$, each satisfying one boundary condition on either half-interval \cite{ito_generators_1996}. Furthermore, the difference of derivatives on the LHS of Eq.\,\eqref{eq:localglobal} can be identified with the Wronskian $W_s(z_\Omega)$ for this same boundary value problem. Altogether this implies the Green's function for a one-dimensional ergodic diffusion is
    \begin{equation}
        G_s(z_0,z_\Omega) = - \frac{p_\mathrm{st}(z_\Omega)}{s}\frac{\langle e^{-s T} \rangle_{z_0}}{\langle e^{-s T} \rangle},
    \end{equation}
    which can be inserted in Eq.\,\eqref{eq:MGFR} to obtain an exact expression for the MGF of $R$ conditional on initializing the diffusion process deterministically at $z_0$,
    \begin{equation}\label{eq:mgfresult}
        \langle e^{-sR} \rangle_{z_0} = \frac{\langle e^{-sT}\rangle_{z_0}}{1 + \frac{s}{\omega_\mathrm{st}}\langle e^{-sT}\rangle}.
    \end{equation}
    This result is equivalent to interpreting $R$ as the functional inverse of the occupation time $\ell_\Omega(t)$ of $\Omega$ evaluated at an $\mathrm{Exp}(\omega_\mathrm{st})$ random threshold, which can be alternatively obtained from the perspective of excursions and inverse local time distributions \cite{pitman_hitting_2003}. Furthermore, if the diffusion process has a characteristic time-scale of the order of the mean hitting time $\langle T \rangle$, we can interpret Eq.\,\eqref{eq:mgfresult} as an interpolation between the two limiting regimes discussed in Section \ref{sec:prem}. If $\langle T \rangle \ll \omega_\mathrm{st}^{-1}$, sampling the trajectory at time points in $C_\Omega$ becomes indistinguishable from sampling the stationary distribution, where setting $T \approx 0$ realizes the Poisson limit of exponential reaction times with intensity $\omega_\mathrm{st}$. Conversely, if $\langle T \rangle \gg \omega_\mathrm{st}^{-1}$, the reaction time is mainly determined by the numerator, which gives the hitting time, so $R \approx T$.\\

    \subsection{Counting statistics and application to biophysical scenarios}\label{sec:counting}
    \color{black}
    Besides providing an exact expression for the reaction time MGF, Eq.\,\eqref{eq:mgfresult} also allows us to obtain the structure of the distribution of $K$ by simply considering $z_0=z_\Omega$, which by the Markov property completely determines the dynamics between two consecutive reactions. With this choice for the initial condition, Eq.\,\eqref{eq:mgfresult} becomes
    Eq.\,\eqref{eq:Kmgf0}
    and $K$, therefore, depends only on properties of the system averaged over its stationary distribution, $\omega_\mathrm{st}$ and $\langle e^{-sT}\rangle$. In this regard, $K$ behaves similarly to the recurrence time of a discrete process \cite{kac_notion_1947}, including having a scale set by the inverse of the stationary distribution.
    
    Moreover, because the MGF of $K$, $\langle e^{-sK}\rangle$, is a function of $\langle e^{-sT} \rangle$, the moments of $K$ can be expressed, by power series composition, as a sum of Bell polynomials over the moments of $T$ \cite{OSullivan_2022}. Due to the factor of $s$ in the denominator, the $n$-th moment of $K$ depends only on the $(n-1)$ first stationary moments of $T$. For the first two moments of $K$, we find
    \begin{align}
        \langle K \rangle &= \omega_{\mathrm{st}}^{-1}\label{eq:meanK}= \langle \mathrm{Exp}(\omega_\mathrm{st})\rangle,\\
        \begin{split}\label{eq:varK} 
        \mathrm{Var}[K] &= \omega_{\mathrm{st}}^{-2} + 2\langle T \rangle \omega_{\mathrm{st}}^{-1} \\
        &= \mathrm{Var}[\mathrm{Exp}(\omega_\mathrm{st})] + 2\langle T \rangle \omega_{\mathrm{st}}^{-1},
        \end{split} 
    \end{align}
    which shows that, on average, $K$ is indistinguishable from the holding times obtained in the Poisson limit, but the variance is always larger (statistically overdispersed) compared to what would be expected from a Poisson process with that rate. 

    As mentioned in section \ref{sec:prem}, $N(t)$ is, due to the Markov property for $\bm z(t)$, a delayed renewal process \cite{ibe_6_2013}, which is fully characterized by the distribution of the holding times $K_i$, given by \eqref{eq:Kmgf0}. Thus, the dependence of $N(t)$ on the underlying trajectories is entirely captured by the stationary distribution and the distribution of hitting times initialized at stationarity.
    
    Many properties of the holding time distribution can only be made explicit by Laplace inversion of the MGF of $K$. We can partially circumvent this by letting the observation time be a random variable $T_{\mathcal O} \sim \mathrm{Exp}(\tau^{-1})$. The scale $\tau^{-1}$ might set a failure rate of the apparatus or decay rate of the trajectory. If we momentarily ignore the delay set by the first hitting time, we can define a new process $\tilde{N}(t)$, such that $\tilde{N}(t) \equiv N(t + T)$. Since $\tilde{N}(T_\mathcal{O}) = n$ whenever $\sum_{i=1}^n K_i \leq T_\mathcal{O} \leq \sum_{i=1}^{n+1} K_i$, and since for any random time $X$, $\mathbb P[X < T_\mathcal{O}] = \langle e^{-X/\tau} \rangle$, we find
    \begin{align}\label{eq:expobservation}
        \mathbb P[\tilde{N}(T_\mathcal{O}) = n] = \langle e^{-K/\tau} \rangle^{n}\left(1 - \langle e^{-K/\tau} \rangle\right).
    \end{align}
    If the initial delay is then included, we simply split into the case where a reaction never occurs and one where the reaction occurs followed by a process identical to $\tilde N(t)$, obtaining,
    \begin{align}
    \begin{split}
        \label{eq:delayrenewalcounts}
        \mathbb P\left[N(T_\mathcal{O}) = n\right] &= \left(1 - \langle e^{-R/\tau}\rangle\right)\delta_{n,0} \\
        &+ \frac{\langle e^{-R/\tau} \rangle^2 \langle e^{-K/\tau} \rangle^{n-1}}{\omega_{\mathrm{st}}\tau}\left(1 - \delta_{n,0}\right).
    \end{split}
    \end{align}
    So far only a random observation time $T_\mathcal{O}$ has been employed, as it circumvents explicit Laplace inversion of the $K_i$ and $T$ distributions. The asymptotic properties of $N(t)$ for a deterministic observation time are nevertheless fixed by the renewal theoretic central limit theorem \cite{smith1958renewal}, meaning $N(t) \approxsim \mathcal N(\langle N \rangle(t), \sigma^2(t))$, where explicit computations for the mean and the variance (see App.\;\ref{app:CLT}) give
    \begin{align}
        \label{eq:cltmean}
        \langle N(t) \rangle &= \omega_\mathrm{st} t, \\
        \label{eq:cltvar}
        \sigma^2(t) &\sim \mathrm{Var}[K]\omega_\mathrm{st}^3 t.
    \end{align}
    Note that we have assumed the initial distribution is stationary. The count is indeed sensitive to initial conditions as, for example, starting directly at the reaction site gives, from \eqref{eq:expobservation},
    \begin{equation}\label{eq:meancount}
        \langle \tilde{N}(T_{\mathcal O}) \rangle = \frac{\omega_{\mathrm{st}} \tau}{\langle e^{-T/\tau} \rangle}.
    \end{equation}
    Ergodicity and the central limit theorem asymptotically constrain what information the counting process contains about the underlying trajectory. We illustrate how these constraints affect inference in two typical cases where detection counts of local motion sensors are used to characterize a biological system. 
    
    \subsubsection{Case study I: activity monitoring illustrates the challenge of deriving transport parameters from counts}
    \noindent Suppose the counts come from a diffusive object inside of a tube of length $d$ that reacts at the midpoint. This can model a protocol for measuring the activity level of fruit flies by counting crossings at some length along a vial \cite{pfeiffenberger_locomotor_2010}. \ed{Up to now, all our results are general for any ergodic Makovian diffusion process. For this particular application, we particularize our results to the simplest model for this situation, which is a Brownian motion with diffusion constant $D$ confined to a one-dimensional domain $x\in[-d/2,d/2]$.}
    For this model, the stationary distribution is uniform and the MGF of the hitting time starting at an initial condition sampled from this uniform probability density function can be calculated analytically for an arbitrary location of the reactive point (see App.\,\ref{app:rbm} for a full calculation of this quantity). 
    
    Under these assumptions for the reaction-diffusion process, the mean number of counts computed from \eqref{eq:expobservation} is, if one starts counting at the first crossing,
    \begin{equation}\label{eq:rbmcount}
        \langle \tilde{N}(T_{\mathcal{O}})\rangle = \sqrt{\frac{\eta^2 \tau}{4 D}}\coth\sqrt{\frac{d^2}{4D\tau}},
    \end{equation}
    If, instead, we consider an arbitrary time to start the observation, such that the particle position is a uniformly distributed random variable within the one-dimensional domain, we get $\langle N(T_{\mathcal{O}}) \rangle = \eta \tau/d$, where the interaction timescale at the reaction domain is $\omega_\mathrm{st}^{-1} = d/\eta$. The mean number of detections decreases monotonically and saturates exponentially with the diffusive timescale, reaching its stationary value when $D\tau \approx d$. This result highlights a perhaps unexpected feature of $N(t)$, namely that the mean count asymptotically gives no information on the underlying motion except for its stationary distribution. Outside this asymptotic regime, it reflects the initial condition that forces the trajectory to dwell near the detector, where counts anticorrelate with the motility parameter, in this case the diffusivity $D$. The anticorrelation effect is only relevant when the observation time is smaller than the typical escape time from the vicinity of the reactive region.
    
\begin{figure}
    \centering
    \includegraphics[width={\linewidth}]{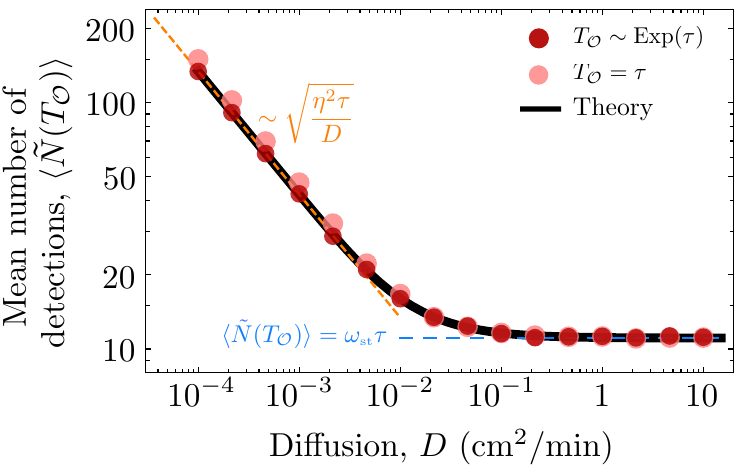}
    \caption{Mean number of detections within a characteristic observation time $\tau$ (log-log scale), conditioned on starting to counts at the first crossing. The symbols correspond to the average over $10^4$ independent realizations of the reaction-diffusion process and the black line to the theoretical prediction in Eq.\,\eqref{eq:rbmcount}. Darker and lighter symbols are obtained with exponentially distributed and constant observation times, respectively. The cyan dashed line shows the Poisson limit $\langle N(T_\mathcal{O})\rangle=\omega_{st}\tau$ and the orange dashed line shows the large-$d$ (or small-$D$) scaling limit in Eq.\,\eqref{eq:rbmcount}. Other parameters: $\eta=10^{-1}$\,cm/s, $d=6.5$\,cm, $\tau=12$\,h, and detector location at the origin $x=0$}\label{fig:compN}
    \end{figure}

    To validate these theoretical results and test how sensitive they are to our assumption of exponentially distributed observation times, we compared the mean number of detections predicted by Eq.\,\eqref{eq:rbmcount} with direct counts performed on numerical simulations of the reaction-diffusion process, considering both exponentially distributed and constant observation times. We parameterized both the reaction-diffusion process and the system domain so they match, at least in orders of magnitude, the typical values one would find in experimental setups monitoring \textit{Drosophila} activity. The results obtained from these numerical simulations show an excellent agreement with the theoretical prediction, and the behavior of the mean number of detections remains qualitatively unchanged when observation times are constant (Fig.\,\ref{fig:compN} and App.\,\ref{app:countRBM} for details on the setup of the numerical simulations). 
    
    \subsubsection{Case study II: ensemble statistics set limits on particle number estimation}\label{subsec:ensemble}
    
    \noindent Counting processes generated by an ensemble of underlying motions occur in questions related to sensing. One classical case is that explored by Berg and Purcell \cite{berg_physics_1977}, which established the precision limits associated with concentration measurements at the cellular scale for chemicals binding to membrane receptors. Subsequent developments have sharpened these estimates by considering effects due to receptor dynamics \cite{bialek_physical_2005, kaizu_berg-purcell_2014} and generalizing to different ambient dimensionalities \cite{bicknell_limits_2015}. Utilizing the theory we have developed so far, we can obtain analogous results for a larger class of drift-diffusion processes.  
    
    As per the previous examples, explicit results for an arbitrary number of particles $M$ will depend on particle lifetimes being i.i.d. $\mathrm{Exp}(\tau^{-1})$ variables, which implies that we are assuming particles degrade at a constant rate $\tau^{-1}$. We further assume that the system starts in its ``well-mixed'' equilibrium distribution. Because we seek an upper bound for the sensing precision, we assume the reactive region acts as perfect detector, meaning we take the continuum limit of $\nu \rightarrow \infty$. Looking at the sum of the observation counts of $M$ such particles (see App.\,\ref{app:occmultipart} for a full calculation), the typical precision of the maximum likelihood estimate of $M$, $\hat{M}$, is
    \begin{align}\label{eq:measurementlimit}
        \frac{\sqrt{\mathrm{Var}[\hat M]}}{M} &= \frac{1}{\sqrt{M}}\sqrt{\frac{2}{\langle e^{-T/\tau} \rangle} - 1 + \mathcal O(M^{-1})} \\ 
        &\sim \frac{1}{\sqrt{M}} \sqrt{1 + \frac{\omega_\mathrm{st}}{\tau}\mathrm{Var}[K]}.
    \end{align}
    Fast motion processes or long observation windows drive the measurement error towards its minimum of pure $M^{-1/2}$ noise, but again the correlation structure of trajectories broadens the distribution.
    
\begin{figure}
    \centering
    \includegraphics[width={\linewidth}]{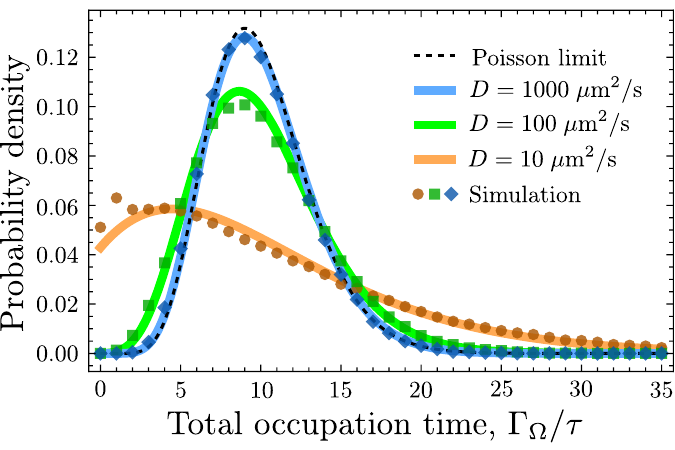}
    \caption{Distribution of the total occupation time w.r.t. $p_{\mathrm{st}}$, $\Gamma_\Omega$, for an ensemble of $M=10$ particles with average lifetime $\tau=10\;\mathrm{s}$ and confined in a spherical domain of radius $\mathcal{R}=10\;\mathrm{\mu m}$. The domain $\Omega$ is defined as a spherical subset of this larger sphere, centered at its origin and with radius $a=1\;\mathrm{\mu m}$. Curves show the theoretical distribution of Eq.\,\eqref{eq:LaplaceInvDist} and symbols are obtained from numerical simulations of the reaction-diffusion process (see App.\,\ref{app:NumMultipart} for details). Different colors correspond to different particle diffusion coefficients, as indicated in the figure legends.}\label{fig:occtime}
    \end{figure}

    This explicitly reproduces the Berg-Purcell limit, considering a spherical detector (``cell'') of radius $a$ centered in a larger sphere of radius $\mathcal{R}$. If particles undergo pure Brownian motion, the local-global correspondence for $L^\dagger = D \bm \nabla^2_{3\mathrm{D}}$ gives an approximate stationary hitting time distribution (see App.\,\ref{app:NumMultipart}),
    \begin{align}\label{eq:3dfptmgf}
        \langle e^{-T/\tau} \rangle &\approx \frac{4\pi Da\tau}{V}\left[1 + f\left(\frac{\mathcal R}{\sqrt{D\tau}}, \frac{a}{\sqrt{D\tau}}\right)\right], \\ \label{eq:auxiliaryfn3d}
        f(x,y) &= y \frac{x\sinh(x-y) - \cosh(x-y)}{x\cosh(x-y) - \sinh(x-y)}.
    \end{align}
    where $V = 4\pi \mathcal{R}^3/3$ is the total volume, and in the regime under consideration $-1 < f(x,y) < y$, with $f$ monotonically increasing in $x$. The theoretical distribution of occupation times predicted from this MGF shows an excellent agreement with numerical simulations of the reaction-diffusion process (Fig.\,\ref{fig:occtime}; see App.\,\ref{app:NumMultipart} for details on these numerical simulations) for a range of diffusivities compatible with biomolecules at the cellular scale. Noting $M/V$ is the particle concentration, Eq.\,\eqref{eq:measurementlimit} reproduces the classical chemosensing limit \cite{berg_physics_1977} as $M, \mathcal R \rightarrow \infty$, ignoring higher order terms in $(Da\tau M/V)^{-1}$,
    \begin{equation}
        \frac{\sqrt{\mathrm{Var}[\hat M]}}{M} \approx \sqrt{\frac{1}{2\pi D a \tau(1 + a/\sqrt{D\tau})M/V}},
    \end{equation}
    where the scale of the observation time is given by the typical particle lifetime, and we note the $a/\sqrt{D\tau}$ term is small by assumption.
    
    \section{Summary and Discussion}\label{sec:conclusion}
    We developed a general framework for analyzing the dynamics of reaction counts in reaction–diffusion processes and applied it to one-dimensional diffusive systems with stationary distributions and point-like reaction domains, where reaction events occur at a constant rate conditioned on occupancy of the reactive region. Under these assumptions, we derived an exact expression for the MGF of the inter-reaction-time distribution as a function of both the Poisson intensity in the reaction domain and the characteristic scales of the diffusion process. \ed{Although biological stochastic dynamics may exhibit non-Markovian or non-ergodic features at short timescales \cite{vilk2022ergodicity,holcman2017100}, Markovian and ergodic diffusion processes provide a standard and accurate description at long times, representing the minimal model consistent with a stationary distribution and providing accurate interpretation of empirical data in cell biology and animal movement \cite{holcman2017100,vilk2022ergodicity,noonan_comprehensive_2019}.}
    
    The structure of this MGF shows how the inter-reaction count process interpolates between reaction- and diffusion-limited regimes. When the reaction intensity is weak or the hitting times are fast ($T\ll K$), consecutive reactions become effectively independent. In this regime, diffusion trajectories may be replaced by sampling from the stationary distribution at a rate set by the reaction intensity and the effective size of the reactive domain, leading to exponentially distributed inter-reaction times. On the contrary, if $T$ contributes to the inter-reaction time significantly, the reaction process departs from Poisson statistics. In this regime, inter-reaction times follow a delayed renewal process, with the properties of the latent stochastic process's trajectories governing the next reaction times.
    
    The Poissonian limit, in which count statistics are determined directly from the stationary distribution of the diffusion process without resolving stochastic trajectories, is important for many biological applications \cite{das_misconceptions_2023}. When the reaction-diffusion process represents the encounter between trajectories of two moving organisms, replacing stochastic trajectories by stationary space-use distributions facilitates the statistical estimation of interactions \citep{noonan_estimating_2021, Fagan2024} and the upscaling of interaction rates to understand how they affect processes such as competition or disease spread \citep{Wilber2022, Menezes2025}. \ed{In disease transmission, this regime yields bounds on transmission rates enabling the analysis of epidemic dynamics using spatial distributions rather than full trajectories. Finally, because stationary space-use distributions are typically easier to compute than trajectory-based statistics, characterizing reaction times through them facilitates extending our results to more complex geometries (provided the diffusive process remains ergodic).}
    
    Beyond forward predictions, the MGF structure also shows that the ergodicity of latent diffusion constrains the inference of movement parameters directly from count statistics. Because the count process is a delayed renewal process, only its mean and variance are asymptotically relevant \cite{smith1958renewal}. We show that mean counts do not reveal any property of the motion process beyond its stationary distribution, whereas the variance does. We put these ideas in more concrete terms by considering two inference scenarios in biophysical systems.
    
    As a first example, we considered a null model for an activity-monitoring assay of a small organism in a test tube \cite{pfeiffenberger_locomotor_2010}, modeled as reflected Brownian motion crossing a barrier. Because this process converges to a uniform stationary distribution, the mean detection count, which might naively be considered a proxy for activity, contains no information about the diffusivity. Inferring motility from a localized detector thus requires the full count distribution.Second, we considered the problem of inferring the size of a population of indistinguishable diffusive objects from the frequency of their interactions with a single detector \cite{rowcliffe_estimating_2008, foster_critique_2012, coblentz_estimating_2021}. We showed that, when the number of particles is large, the achievable inference precision is bounded by the variance of the inter-reaction time distribution, recovering and generalizing the classic Berg-Purcell limit to a much broader class of diffusions. One relevant application of these results is estimating animal abundance from camera-trap detections \cite{Gilbert2021}, while accounting for realistic movement patterns that may be non-Brownian and landscape-dependent, but are accurately described by ergodic Markovian processes at long timescales \cite{noonan_search_2023,noonan_comprehensive_2019}.
    
    In this second example, spherical symmetry reduces a three-dimensional geometry to an effectively one-dimensional problem. Generalizing Eq.\,\eqref{eq:Kmgf0} to higher dimensions may be possible under two conditions. First, the diffusion operator must be expressible in the Sturm-Liouville form of Eq.\,\eqref{eq:stationaryoperator}. This is possible for some choices of the stochastic process, for example considering a gradient drift with additive noise, where the stationary measure is a Boltzmann distribution. Second, the local–global correspondence between near-boundary and equilibrium-initialized hitting times needs to be expressed as a surface integral over $\partial\Omega$. Linking this representation to the Green’s function requires an additional assumption of local isotropy near the reactive domain. \ed{More complex geometries, such as heterogeneous landscapes or networks, may not admit analytical treatment beyond the reaction-limited regime discussed above. However, by statistically estimating occupation probabilities and typical mean hitting times, we can still infer the distribution of detection events.}
    
    In the two biophysical applications discussed above, we modeled the diffusion process as purely Brownian motion and used reflecting boundary conditions to ensure stationarity. Markovian processes in which stationarity is ensured by confining drifts have several applications across many disciplines. For example, the Ornstein-Uhlenbeck (OU) process, which combines a linear restoring force with Brownian noise, accurately describes the long-time behavior of animal-tracking data across multiple datasets \cite{noonan_comprehensive_2019}. \ed{Because OU first-hitting times and stationary distributions can be calculated analytically, one immediate application of this framework is to study wildlife-vehicle collisions in range-resident species \cite{GarciaDeFigueiredo2025}. In such systems, collision events constitute a counting process generated by range-resident trajectories, matching directly the modeling assumptions developed here.}
    
    \ed{Likewise, interaction between many individuals within a landscape may be understood in a similar manner by considering the joint process of trajectories in a higher-dimensional phase space. In this description, pairwise interactions are characterized by the times when two trajectories come sufficiently close. These proximity events can be represented as visits to specific \textit{interaction regions} in the high-dimensional phase space of their positions \cite{sugaya_analysis_2018}. The resulting pattern of pairwise encounters can then be used to improve population-level models \cite{Menezes2025}. While standard coarse-grained reaction models assume that events occur after exponentially distributed waiting times, allowing more general (non-exponentially) inter-reaction waiting-time distributions provides a simple way to capture realistic deviations from this assumption \cite{Castro2018}.
    
    Situations where trajectories are non-Markovian represent a natural extension of our framework, because techniques for studying the full distribution of first-hitting times are still being developed \cite{marrisPersistentAntipersistentMotion2024a}. When additional internal variables $B$ (e.g., velocity or conformational state) are included, the dynamics are Markovian only in the extended state space $(x,B)$ and the reactive region becomes $\Omega \times B$. If these internal variables have finite correlation times, reaction events depend on the joint state of the system, and the resulting reaction process is expected, in general, to lose the renewal structure that enables the asymptotic characterization obtained here. In backward inference problems, autocorrelated velocities, such as those arising from active transport, can improve classical chemosensing limits \cite{godecSignalFocusingActive2015, smogreGettingCellPhysical2020}, suggesting that non-Markovian latent processes could likewise provide interesting corrections to our inference results.
    
    Related challenges arise when the ergodicity of the underlying process is relaxed. For autonomous Markovian processes, the two remaining possibilities are null-recurrence (non-existence of $p_\mathrm{st}$ and transience ($T \rightarrow \infty$ with finite probability). The null-recurrent case can be treated with techniques similar to those employed here by considering a non-normalizable stationary measure instead of a limiting probability distribution, although this does not lead to the class of limiting Poisson point processes discussed in the present framework \cite{pitman_hitting_2003}. Therefore, both in non-Markovian and non-ergodic regimes, the asymptotic renewal properties underlying the reaction statistics may be lost.
    
    Similar issues may arise in non-autonomous scenarios with time-dependent parameters or finite observation windows. In biological applications, for example, parameters are often time-dependent due to both exogenous and endogenous factors, and many processes have finite lifetimes, which can impact the asymptotic counting regime. Extending our results to these cases is another important direction for future work, but it presents some challenges because our approach relies on frequency-space methods developed for autonomous systems. While finite lifetimes can break ergodicity by imposing a time cut-off, situations of population turnover near a steady state could be treated by allowing the processes to reset periodically \cite{montero_continuous-time_2017}. This would lead to a possible distinction between the stationary distribution of a single process and the initialization distribution of the population in the count statistics, both of which are well motivated by biological phenomena.}
    
    \section*{Acknowledgments}
    We thank William Bialek and Eliezer D. Gurarie for insightful discussions at different stages of the study and Luisa Ramirez for valuable feedback on the manuscript. This work was partially funded by the Center of Advanced Systems Understanding (CASUS), which is financed by Germany’s Federal Ministry of Research, Technology, and Space (BMFTR) and by the Saxon Ministry for Science, Culture and Tourism (SMWK) with tax funds on the basis of the budget approved by the Saxon State Parliament. RMG and BGF were partially supported by Instituto Serrapilheira through grant Serra-1911-31200; the Simons Foundation through grant 284558FY19; and FAPESP through a BIOTA Jovem Pesquisador Grant 2019/05523-8 (RMG), ICTP-SAIFR 2021/14335-0 (RMG), and a Master's fellowship 2019/26736-0 (BGF). 
    
\providecommand{\noopsort}[1]{}\providecommand{\singleletter}[1]{#1}%
%

    
    \onecolumngrid
    \appendix
    \section*{Appendices}
    
    \section{Derivation of the backwards Fokker-Planck equation for the moment generating function}\label{app:MGFderivation}
    
    In the adjoint Fokker-Planck equation \eqref{eq:FPEback} the variable $\bm z$ can be integrated over. This operation defines the survival function $S(t|\bm z_0)$, which is the complementary cumulative distribution function of $R$,
    \begin{equation}
        S(t|\bm z_0) = \int P(\bm z,t|\bm z_0)\mathrm{d}\bm z,
    \end{equation}
    and satisfies the same backwards equation as $P(\bm z,t|\bm z_0)$, with initial condition $S(0|\bm z_0) = 1$. Recurrence furthermore imposes $S(t \rightarrow \infty|\bm z_0) = 0$. 
    The problem of determining the statistics of $R$ can thus be reduced to solving the equation
    \begin{equation}\label{eq:survival}
        \frac{\partial S(t|\bm z_0)}{\partial t} = \left[\hat L^\dagger_{\bm z_0} - \omega(\bm z_0)\right]S(t|\bm z_0).
    \end{equation}
    The survival equation \eqref{eq:survival} can be equivalently written as an equation for the moment generating function (MGF) of $R$, $\langle e^{-sR}\rangle_{\bm z_0}$, which coincides with the Laplace transform of the probability density function $\phi(t|\bm z_0) = -\partial S/{\partial t}$. 
    
    We thus conclude that, in a first approximation of $\omega$ as a delta function that ignores the shape of the small region $\Omega$, the MGF of $R$ is a Green's function for the operator $\left[\hat L^\dagger_{\bm z_0} - s\right]$ perturbed by a delta function. This means Eq.\,\eqref{eq:MGF} may be pre-multiplied by the Green's function of $\left[\hat L^\dagger_{\bm z_0} - s\right]$, $G_s(\bm z_0,\bm y_0)$, and integrated over to find,
    \begin{equation}
        \langle e^{-sR} \rangle_{\bm{z}_0} = \eta(\langle e^{-sR} \rangle_{\bm{z}_\Omega} - 1) G_s(\bm{z}_0,\bm{z}_\Omega).
    \end{equation}
    Substituting $\bm z_0 = \bm z_\Omega$ and solving for $\langle e^{-sR} \rangle_{\bm{z}_\Omega}$ gives \eqref{eq:MGFR}.
    
    \section{Count distributions for deterministic observation times}\label{app:CLT}
    
    \noindent If the observation time is a fixed number $t$, the distribution of occupation probabilities involves a Laplace inversion, as, in the language of Eq.\;\eqref{eq:expobservation},
    \begin{equation}
        \mathbb P[N(T_\mathcal O) = n] = \frac{1}{\tau} \int_{0}^\infty \mathbb P[N(t) = n]e^{-t/\tau}\mathrm{d}t.
    \end{equation}
    For the delayed renewal process, this distribution is, using Eq.\;\eqref{eq:delayrenewalcounts},
    \begin{equation}
        \mathbb P[N(t) = n] = \frac{1}{2\pi i}\int_{\alpha - i\infty}^{\alpha + i\infty} \left[ \left(1 - \langle e^{-sR} \rangle\right) \delta_{n,0} + \frac{s}{\omega_{\mathrm{st}}}\langle e^{-sR} \rangle^2 \langle e^{-sK}\rangle^{n-1}\left(1 - \delta_{n,0}\right) \right] \frac{e^{st}}{s}\mathrm d s.
    \end{equation}
    Evaluating the moments of this distribution, we find that the mean has a universal form that does not depend on the hitting time distribution, 
    \begin{equation}\label{eq:delaymeancount}
        \langle N(t) \rangle = \frac{1}{2\pi i}\int_{\alpha - i\infty}^{\alpha + i\infty} \frac{1}{\omega_{\mathrm{st}}}\left(\frac{\langle e^{-sR} \rangle}{1 - \langle e^{-sK} \rangle}\right)^2e^{st}\mathrm d s = \frac{\omega_{\mathrm{st}}}{2\pi i}\int_{\alpha - i\infty}^{\alpha + i\infty}\frac{e^{st}}{s^2}\mathrm d s = \omega_{\mathrm{st}} t,
    \end{equation}
    whereas the second moment follows
    \begin{equation}\label{eq:Asecond}
        \langle N(t)^2 \rangle = \frac{\omega_\mathrm{st}^2}{2\pi i}\int_{\alpha - i\infty}^{\alpha + i\infty} \frac{1 + \langle e^{-sK} \rangle}{\langle e^{-sR}\rangle}\frac{e^{st}}{s^3}\mathrm d s.
    \end{equation}
    Evaluating Eq.\,\eqref{eq:Asecond} in general depends on details of the diffusion, namely on the complex-analytic structure of the moment generating functions. However, the long-time asymptotics are determined by the singularity at $s = 0$, and we may expand the integrand using
    \begin{equation}
        \frac{1 + \langle e^{-sK} \rangle}{\langle e^{-sR}\rangle} = \left[2 - s\langle K \rangle + \frac{s^2}{2}\langle K^2 \rangle + \mathcal O(s^3)\right]\left[1 - s \langle R \rangle + \frac{s^2}{2}\left(2 \langle R\rangle^2 - \langle R^2 \rangle\right) + \mathcal O(s^3)\right],
    \end{equation}
    to compute
    \begin{equation}
        \mathrm{Res}_{s = 0} \frac{1 + \langle e^{-sK} \rangle}{\langle e^{-sR}\rangle}\frac{e^{st}}{s^3} = \omega_{\mathrm{st}}^2 \left[\langle T\rangle^2 - \mathrm{Var}[T] - \mathrm{Var}[K] + \mathrm{Var}[K] \omega_\mathrm{st} t + t^2\right].
    \end{equation}
    This recapitulates the central limit theorem, Eq.\,\eqref{eq:cltvar}, where
    \begin{equation}
        \mathrm{Var} [N(t)] \sim \mathrm{Var}[K]\omega_{\mathrm{st}}^3 t.
    \end{equation}
    \section{Distributions for Reflected Brownian Motion in 1D}\label{app:rbm}
    
    \noindent One-dimensional Brownian motion is defined by
    \begin{align}
        \label{eq:browniandrift}
        \mu(z) &= 0,\\
        \label{eq:brownianvar}
        \Sigma(z)  &= \sqrt{2D},
    \end{align}
    where $D$ is the diffusion constant. The adjoint Fokker-Planck operator from Eq.\;\eqref{eq:fpoperator} is simply 
    \begin{equation}
        \hat L^{\dagger}_z = D \frac{\mathrm{d}^2}{\mathrm{d}z^2}.
    \end{equation}
    This motion has a stationary distribution when limited to a box of size $d$, parameterized by the interval $[-d/2, d/2]$ with reflecting boundary conditions, or Neumann boundary conditions for $\hat L$ at $x = \pm d/2$. The stationary distribution 
    $p_\text{st}$ satisfies $\hat L p_\mathrm{st} = 0$, so $p_\mathrm{st}$ is an affine function with $p_\mathrm{st}'(\pm d/2) = 0$ and thus a uniform distribution
    \begin{equation}\label{eq:rbmstationary}
        p_{\mathrm{st}}(x) = d^{-1}.
    \end{equation}
    The MGF of the hitting time $T$ to a point $z_\Omega$, starting from an initial position $z_0$, satisfies Eq.\;\eqref{eq:MGF} in the limit of infinite reaction rate, namely
    \begin{equation}
        \left(\frac{\mathrm{d}^2}{\mathrm{d}z_0^2} - \frac{s}{D}\right) \langle e^{-sT} \rangle_{z_0} = 0.
    \end{equation}
    This equation admits as fundamental solutions
    \begin{equation}
        \psi_{\pm}(z_0) = \exp\left({\sqrt{\frac{s}{D}} \left(\pm d - z_0\right)}\right) + \exp\left({\sqrt{\frac{s}{D}} z_0}\right),
    \end{equation}
    which satisfy the right ($+$) and left ($-$) Neumann boundary conditions respectively, and are monotonic in the interval. The MGF must evaluate to unity at $z_\Omega$, since the hitting time is zero for that initial condition, so the unique solution can be written as
    \begin{equation}\label{eq:fptgeneralsol}
    \langle e^{-sT} \rangle_{z_0} =
        \begin{cases}
        \psi_{-}(z_0)/\psi_{-}(z_\Omega) & z_0 \leq z_\Omega, \\
        \psi_+(z_0)/\psi_+(z_\Omega) & z_0 > z_\Omega.
        \end{cases}
    \end{equation}
    Using the above result with the local-global correspondence of Eq.\;\eqref{eq:localglobal}, we find the MGF of the stationary hitting time to be
    \begin{equation}
        \langle e^{-sT} \rangle = \sqrt{\frac{D}{d^2 s}}\sinh\left[\sqrt{\frac{s}{D}}d\right]\text{sech}\left[\sqrt{\frac{s}{D}}\left(\frac{d}{2}-z_\Omega\right)\right]\text{sech}\left[\sqrt{\frac{s}{D}}\left(\frac{d}{2}+z_\Omega\right)\right],
    \end{equation}
    which when specialized to $z_\Omega=0$ in Eq.\;\eqref{eq:meancount} gives the mean count $\langle N(T_\mathcal{O}) \rangle$ in Eq.\;\eqref{eq:rbmcount}. 
    The stationary mean hitting time is
    \begin{align}
        \langle T \rangle &= - \frac{\partial}{\partial s}\bigg |_{s = 0^+} \langle e^{-sT} \rangle = \frac{d^2}{12D} + \frac{z_\Omega^2}{D},
    \end{align}
    which may be used together with Eqs.\;\eqref{eq:rbmstationary} and \eqref{eq:varK} to give $\mathrm{Var}[K]$.
    
    \section{Numerical simulations of one-dimensional reflected Brownian motion with a punctual reaction domain}\label{app:countRBM}
    
    We tested the accuracy of the theoretical prediction for the mean number of counts within an observation time $T_{\mathcal{O}}$ by performing direct numerical simulations of a one-dimensional reflected Brownian motion, $z \in [-d/2,\,d/2]$, interacting with a punctual reaction domain located at $d=0$. For every realization of this counting process, we generated an exponentially distributed observation time $T_\mathcal{O}\sim \mathrm{Exp}(\tau)$ and generated a realization of the reflected Brownian motion, $z(t)$ with $t\in[0, T_\mathcal{O}]$ using a standard Euler-Maruyama algorithm \cite{Toral2014} with time step $\Delta t$. Since our theoretical results are conditional on ignoring the initial interaction between the Brownian trajectory and the reaction domain, we set $z_0=d=0$. 
    
    Simulating the interaction process between the reaction domain and the Brownian trajectory is challenging because we are considering that $\Omega$ is punctual and hence the probability that $z(t)$ lands there is zero. We overcame this issue considering that the Brownian trajectory interacts with the reaction domain with a probability $P_\times<1$ every time it crosses the origin. A key part of the simulation setup is thus to relate this reaction probability $P_\times$ to $\eta$. Establishing this relationship is not trivial because Brownian trajectories are non-differentiable and the number of times the boundary at $d$ is crossed diverges to infinity in the continuum-time limit. Rigorous treatment of these issues \cite{singer_brownian_2005,singer_partially_2007,erban_reactive_2007} requires thus taking into account that the limit of $\Delta t \rightarrow 0$ is singular when reactive boundaries are present. One must use the method of matched asymptotic expansions in $\Delta t \rightarrow 0$, where a boundary-layer of width proportional to $\sqrt{\Delta t}$ is present close to $z = d$. We follow these analyses to derive a relation between $P_\times$ and $\eta$ in analogy with partially reflecting boundaries \citep{erban_reactive_2007},
    \begin{equation}
        P_\times = \sqrt{\frac{\pi \Delta t}{4D}}\eta.
    \end{equation}

    \section{Occupation time distributions for multi-particle systems}\label{app:occmultipart}
    In the $\nu \rightarrow \infty$ limit, the reaction count diverges, and the natural object of study becomes the proportion of time spent by the $i$-th trajectory $i$ at the reactive region, which defines the occupation time up to time $t$ with respect to the stationary distribution, $\gamma_i(t) \equiv N_i(t)/\omega_{\mathrm{st}}$ as $\omega_{\mathrm{st}}\tau \rightarrow \infty$. Note that this quantity differs from the occupation time by a factor of the total stationary occupation probability of $\Omega$, that is, $\ell_i(t) = \gamma_i(t)\int_\Omega p_\mathrm{st}(z)\mathrm dz$. From Eqs.\,\eqref{eq:delayrenewalcounts} and \eqref{eq:mgfresult}, it follows
    \begin{align}
    \begin{split}\label{eq:occupationdist}
        \mathbb P[\gamma_i(T_\mathcal O) \in \text{d}t] &= \left(1 - \langle e^{-T/\tau} \rangle\right)\delta(t)\text{d}t \\
        &+ \langle e^{-T/\tau} \rangle^2 \exp\left(-\langle e^{-T/\tau}\rangle \frac{t}{\tau}\right)\frac{\text{d}t}{\tau},
    \end{split}
    \end{align}
    which implies
    \begin{align}
        \label{eq:meanoccupation}
        \langle \gamma_i(T_\mathcal{O}) \rangle &= \tau, \\
        \label{eq:occvariance}
        \mathrm{Var}\left[\gamma_i(T_\mathcal{O})\right] &= \left(\frac{2}{\langle e^{-T/\tau} \rangle} - 1\right) \tau^2.
    \end{align}
    In the Poisson process limit, $\gamma_i(T_\mathcal{O}) \approxsim \mathrm{Exp}(\tau^{-1})$, so again the average number of counts is identical to what would be expected from a process at a constant rate, with the variance being larger.
    
    The distribution of the occupation time of a particle in Eq.\;\eqref{eq:occupationdist} can be equivalently written as a MGF
    \begin{equation}\label{eq:appDmgf}
        \langle e^{-s \gamma} \rangle = 1 - \langle e^{-T/\tau} \rangle +  \frac{\langle e^{-T/\tau} \rangle^2}{\tau s + \langle e^{-T/\tau} \rangle}.
    \end{equation}
    
    \noindent Next, because the MGF of a sum of independent variables is the product of the individual MGFs, the distribution for $\gamma/\tau$ can be found by Laplace inversion  of \eqref{eq:appDmgf} in $s$. For $t > 0$, the PDF $p_{M}$ is
    \begin{align}
    \begin{split}\label{eq:mlederivation}
         p_{M}(t)|_{t > 0} &= \frac{1}{2\pi i}\int_{\alpha - i\infty}^{\alpha + i\infty} \langle e^{-s\gamma} \rangle^M e^{s t}\text{d}s = \mathrm{Res}_{\tau s = -\langle e^{-T/\tau}\rangle} \langle e^{-s\gamma} \rangle^M e^{s t} \\
        &= \frac{1}{(M-1)!} \frac{\text{d}^{M-1}}{\text{d} s^{M-1}}\bigg|_{\tau s = -\langle e^{-T/\tau}\rangle} \left(\tau s + \langle e^{-T/\tau} \rangle\right)^M\left\{1 - \langle e^{-T/\tau} \rangle +  \frac{\langle e^{-T/\tau} \rangle^2}{\tau s + \langle e^{-T/\tau} \rangle}\right\}^M e^{st} \\
        &= \sum_{k=1}^M \binom{M}{k} \frac{(1-\langle e^{-T/\tau}\rangle)^{M-k}\langle e^{-T/\tau}\rangle^{2k}}{(k-1)!}\frac{t^{k-1}}{\tau^k}e^{-\langle e^{-T/\tau}\rangle t/\tau} \\&= \frac{1}{\tau} \left(1-\langle e^{-T/\tau}\rangle\right)^{M-1}\langle e^{-T/\tau}\rangle^2 L_{M-1}^{(1)}\left(-\frac{\langle e^{-T/\tau}\rangle^2}{1-\langle e^{-T/\tau}\rangle} \frac{t}{\tau}\right)e^{-\langle e^{-T/\tau}\rangle t/\tau},
    \end{split}
    \end{align}
    where $L_{n}^{(
    \beta
    )}$ denotes an associated Laguerre polynomial. The last identity in \eqref{eq:mlederivation} comes from the fact that the ratio of successive terms in the sum is a rational function of degree 1, so it can be expressed as the hypergeometric function $_1F_1$ with negative integer parameter, which is always an associated Laguerre polynomial \cite{noauthor_dlmf_nodate}. Including the term for $t = 0$, the distribution \eqref{eq:mlederivation} is
    \begin{equation}\label{eq:LaplaceInvDist}
        p_M(t) = \left(1-\langle e^{-T/\tau}\rangle\right)^{M}\delta(t)+ \frac{1}{\tau} \left(1-\langle e^{-T/\tau}\rangle\right)^{M-1}\langle e^{-T/\tau}\rangle^2 L_{M-1}^{(1)}\left(-\frac{\langle e^{-T/\tau}\rangle^2}{1-\langle e^{-T/\tau}\rangle} \frac{t}{\tau}\right)e^{-\langle e^{-T/\tau}\rangle t/\tau}.
    \end{equation}
    A maximum likelihood estimate $\hat M$ of $M$, given a non-zero volume of detections, follows $\mathbb P[\hat M = m|\Gamma_\Omega = t] \propto p_M(t)$. Luckily, this distribution is easily normalized over $M = 1, 2, 3...$, since looking at the first step in Eq.\;\eqref{eq:mlederivation}
    \begin{align}
        \sum_{M=1}^\infty p_M(t) &= \frac{1}{2\pi i}\int_{\alpha - i\infty}^{\alpha + i\infty} \frac{\langle e^{-s\gamma} \rangle}{1 - \langle e^{-s\gamma} \rangle} e^{s t}\text{d}s = \frac{1}{2\pi i\tau}\int_{\alpha - i\infty}^{\alpha + i\infty} \frac{e^{st}}{s}\mathrm{d}s = \tau^{-1}, \\
        \implies \mathbb P[\hat M = m|\Gamma_\Omega] &= (1-\langle e^{-T/\tau} \rangle)^{m-1}\langle e^{-T/\tau} \rangle^2 L_{m-1}^{(1)}\left(-\frac{\langle e^{-T/\tau} \rangle^2}{1-\langle e^{-T/\tau} \rangle} \frac{\Gamma_\Omega}{\tau}\right) e^{-\langle e^{-T/\tau}\rangle \Gamma_\Omega/\tau}.
    \end{align}
    Similarly, we can evaluate the moments of $\hat M$ by computing the residue of a geometric series
    \begin{align}
    \begin{split}
        \langle x^{\hat M} \rangle &= \frac{\tau}{2\pi i}\int_{\alpha - i\infty}^{\alpha + i\infty} \frac{\langle e^{-T/\tau}\rangle + \tau s}{(1-x)\langle e^{-T/\tau}\rangle + \left[1 - (1-\langle e^{-T/\tau}\rangle) x\right]\tau s}e^{s\Gamma_\Omega}\mathrm{d}s \\ &= \frac{\langle e^{-T/\tau}\rangle^2}{\left[1 - (1-\langle e^{-T/\tau}\rangle) x\right]^2}\exp\left(-\frac{\Gamma_\Omega}{\tau}\frac{(1-x)\langle e^{-T/\tau}\rangle}{1 - (1 - \langle e^{-T/\tau}\rangle) x}\right),
    \end{split}\\
        \langle \hat M \rangle &= \left(x \frac{\mathrm{d}}{\mathrm{d}x}\right)\bigg |_{x = 1}\langle x^{\hat M} \rangle = 2\left(\frac{1}{\langle e^{-T/\tau}\rangle} - 1\right) + \frac{\Gamma_\Omega}{\tau}, \\
        \mathrm{Var}[\hat M] &= \left(x \frac{\mathrm{d}}{\mathrm{d}x}\right)^2\bigg|_{x = 1}\langle x^{\hat M} \rangle - \langle \hat M \rangle^2 = \frac{2}{\langle e^{-T/\tau}\rangle}\left(\frac{1}{\langle e^{-T/\tau}\rangle} - 1\right) + \left(\frac{2}{\langle e^{-T/\tau}\rangle} - 1\right)\frac{\Gamma_\Omega}{\tau}.
    \end{align}
    The estimate in Eq.\;\eqref{eq:measurementlimit} is obtained by evaluating the equation for $\mathrm{Var}[\hat M]$ at the typical value of $\Gamma_\Omega \approx M\tau$.
    
    \section{Calculations and numerical simulations of the multi-particle case study}\label{app:NumMultipart}
    
    \noindent Brownian motion in three dimensions follows Eqs.\;\eqref{eq:browniandrift} and \eqref{eq:brownianvar} independently in each of the three Cartesian coordinates. Equivalently, it has a (self-adjoint) Fokker-Planck operator $L^\dagger = D\bm{\nabla}^2$. Since the domain under consideration is spherical, the system may equivalently be described by its radial coordinate $r(t) = \|\bm z(t) \|$, which is a Bessel process 
    \begin{align}
        \mu(r)    &= \frac{2D}{r}, \\
        \Sigma(r) &= \sqrt{2D},
    \end{align}
    or in terms of the equation for the first-passage time,
    \begin{equation}
        \left[\frac{2D}{r_0}\frac{\partial}{\partial r_0} + D \frac{\partial^2}{\partial r_0^2} - s\right]\langle e^{-sT} \rangle_{r_0} = 0,
    \end{equation}
    whereas the equation for the reaction time is
    \begin{equation}
        \left[\frac{2D}{r_0}\frac{\partial}{\partial r_0} + D \frac{\partial^2}{\partial r_0^2} - s - \nu \bm 1[r_0 < a]\right]\langle e^{-sR} \rangle_{r_0} = -\nu\bm 1[r_0 < a].
    \end{equation}
    The fundamental solutions obeying the reflecting boundaries at $r = 0$ and $r = \mathcal R$ can be taken to be
    \begin{align}
        \psi_-(r_0) &= \frac{\mathcal R}{r_0} \sinh \left(\sqrt{\frac{s}{D}}r_0\right), \\
        \psi_+(r_0) &= \frac{\mathcal R}{r_0} \left[\exp\left(\sqrt{\frac{s}{D}}(r_0-2\mathcal R)\right)-\frac{1 - \sqrt{\mathcal R^2 s/D}}{1 + \sqrt{\mathcal R^2 s/D}}\exp\left(-\sqrt{\frac{s}{D}}r_0\right) \right],
    \end{align}
    but now we must note that Eq.\;\eqref{eq:fptgeneralsol} holds only for $r > a$, since otherwise we want to impose $T = 0$. This is where we employ our results as an approximation, since rigorously they would only hold for a reactive spherical shell of radius $a$, not a solid sphere. It is nevertheless true that
    \begin{equation}\label{eq:besselstationaryfpt}
        \langle e^{-sT} \rangle = \int_0^a p_\mathrm{st}(r_0)\mathrm{d}r_0 + \int_{a}^\mathcal R p_\mathrm{st}(r_0) \langle e^{-sT}\rangle_{r_0} \mathrm{d}r_0,
    \end{equation}
    So our approximation is effectively that, since $a \ll \mathcal R$, the spatial structure of the solution for $r < a$ may be ignored. The stationary distribution has $p_\mathrm{st}(r) \propto r^2$, so the probability of having the system initalized inside the reactive domain is $(a/\mathcal R)^3$. Thus, in Eq.\;\eqref{eq:besselstationaryfpt} the first term is ignored, whereas for the second term may be evaluated exactly using a one-sided local-global correspondence,
    \begin{equation}
        \langle e^{-s T}\rangle \approx -\frac{4 \pi D a^2}{Vs}\frac{\mathrm{d}}{\mathrm{d}r_0}\frac{\psi_+(r_0)}{\psi_+(a)}\bigg |_{r_0 = a},
    \end{equation}
    which gives the result in Eq.\;\eqref{eq:3dfptmgf}. Similarly, the distribution of the time $K$ is approximately given by Eq.\;\eqref{eq:Kmgf0}.
    
    As detailed in Supplement \ref{app:countRBM}, we performed simulations of the Bessel process by integrating the equation for $r(t) \in [0,\mathcal{R}]$ using an Euler-Maruyama method with a timestep $\Delta t$. We sampled each particle lifetime as an independent $\mathrm{Exp}(\tau^{-1})$ random variable and the initial positions from the stationary distribution. The occupation time for each particle is computed by summing the number of timesteps $\Delta t$ for which $r(t) < a$, and these are subsequently aggregated into the total occupation time.
    
    \end{document}